\documentclass[manuscript,screen,nonacm]{acmart}

\usepackage{array}
\usepackage{booktabs}
\usepackage{float}  
\usepackage{graphicx}  
\usepackage{hyperref} 

\usepackage{algorithm}
\usepackage{algorithmic}
\usepackage{xcolor}
\usepackage{listings}

\usepackage{tcolorbox}
\tcbuselibrary{breakable}

\definecolor{codegreen}{rgb}{0,0.6,0}
\definecolor{codegray}{rgb}{0.5,0.5,0.5}
\definecolor{codepurple}{rgb}{0.58,0,0.82}

\lstdefinestyle{mystyle}{
    commentstyle=\color{codegreen},
    keywordstyle=\color{magenta},
    numberstyle=\tiny\color{codegray},
    stringstyle=\color{codepurple},
    basicstyle=\ttfamily\footnotesize,
    breakatwhitespace=false,         
    breaklines=true,                 
    captionpos=b,                    
    keepspaces=true,                 
    numbersep=5pt,                  
    showspaces=false,                
    showstringspaces=false,
    showtabs=false,                  
    tabsize=2
}

\begin{document}


\title{Apple vs. Oranges: Evaluating the Apple Silicon M-Series SoCs for HPC Performance and Efficiency}

\author{Paul Hübner}
\author{Andong Hu}
\author{Ivy Peng}
\author{Stefano Markidis}

\affiliation{
  \institution{KTH Royal Institute of Technology}
  \city{Stockholm}
  \country{Sweden}
}

\begin{abstract}
This paper investigates the architectural features and performance potential of the Apple Silicon M-Series SoCs (M1, M2, M3, and M4) for HPC. We provide a detailed review of the CPU and GPU designs, the unified memory architecture, and coprocessors such as Advanced Matrix Extensions (AMX). We design and develop benchmarks in the Metal Shading Language and Objective-C++ to assess FP32 computational and memory performance. We also measure power consumption and efficiency using Apple's \texttt{powermetrics} tool. Our results show that the M-Series chips offer up to 100~GB/s memory bandwidth, and significant generational improvements in computational performance, with up to 2.9 FP32 TFLOPS on the M4. Power consumption varies from a few Watts to 10-20 Watts, with more than 200 GFLOPS per Watt efficiency of GPU and accelerator reached by all four chips. Despite limitations in FP64 support on the GPU, the M-Series chips demonstrate strong potential for energy-efficient HPC applications. While existing HPC solutions such as the Nvidia Grace-Hopper superchip outperform Apple Silicon in both memory bandwidth and computational performance, we see that the M-Series provides a competitive power-efficient alternative to traditional HPC architectures and represents a distinct category altogether -- forming an \textit{apples-to-oranges} comparison.
\end{abstract}


\maketitle

\section{Introduction} 
Systems on a Chip (SoCs) have emerged as viable candidates for HPC deployments. An SoC integrates multiple components -- such as the central processing unit (CPU), graphics processing unit (GPU), memory, storage controllers, and specialized accelerators -- into a single chip. Different from traditional HPC architectures, which rely on discrete components connected via motherboard interconnects (e.g., CPUs, GPUs, memory, and networking cards), SoCs provide these functionalities within a unified architecture. This tight integration reduces interconnect latency, power consumption, and system footprint, which are critical factors in HPC~\cite{rajovic2013supercomputing,selinger2016evaluation}.  

An advantage of SoCs for HPC is their unified memory architecture. This eliminates the overhead of data transfers between separate processing units. In addition, the tight integration of components enables more efficient power management, reducing energy consumption while maintaining computational performance. This power efficiency is particularly relevant for large-scale HPC deployments, where energy costs and thermal constraints are of particular consideration. Workloads exploiting the tight coupling of compute, memory, and accelerators can benefit from improved efficiency and reduced operational costs.  

The Apple Silicon M-Series, including the M1, M2, M3, and M4 chips, is a prime example of SoC architecture that integrates multiple computational components. These chips feature ARM-based performance and efficiency cores, along with specialized coprocessors such as Advanced Matrix Extensions (AMX), for matrix operations, integrated GPUs, and a Neural Engine. The unified memory architecture, shared between the CPU, GPU, and accelerators, aims to increase bandwidth and reduce memory latency.

The M-Series CPUs support a wide range of numerical precisions, including double-precision (FP64), single-precision (FP32), and lower-precision formats, making them feasible for HPC workloads requiring high numerical accuracy. In contrast, the M-Series GPUs lack native FP64 support (which can be emulated) but are optimized for FP32 and mixed-precision (FP16/INT8) computations. While FP32 remains viable for many HPC applications, numerical accuracy must be carefully evaluated depending on workload requirements. In this study, we assess the M-Series GPUs' suitability for HPC workloads by focusing on their computational efficiency in FP32.  

Although initially designed for consumer devices like laptops and desktops, the potential of Apple Silicon M-Series SoCs -- comprising the M1, M2, M3, and M4 -- for HPC is an open question, especially given their reputation as extremely power-efficient chips. In other words, it remains to be seen whether Apple's M-Series chips can compete with traditional architectures or if comparing them is simply an \textit{apples-to-oranges} exercise. In this work, we provide a detailed architectural review of the Apple Silicon M-Series, focusing on key features such as the CPU and GPU designs, unified memory, and the AMX units, while also discussing their programming models and potential for HPC workloads. This paper further investigates whether the Apple Silicon M-Series chips are a viable approach for HPC workloads by assessing three aspects -- data movement, computational tasks, and energy consumption -- to determine whether they meet the demands of computationally intensive workloads in professional and scientific computing. To evaluate performance, we design and implement two benchmarks: a STREAM (for both CPU and GPU)~\cite{mccalpin1995stream} and \texttt{GEMM} (matrix-matrix multiply) with Metal Shading Language and Objective-C++. In addition, we develop a framework to measure the computational intensity and power efficiency via Apple's \texttt{powermetrics} utility. All computations are performed in single-precision (FP32), to compare Apple Silicon M-Series CPU and GPU performance. 

Our results show that both CPU and GPU can reach memory bandwidth comparable to that of unified memory On M4, a close to the theoretical peak bandwidth of 100 GB/s can be obtained. In terms of FP32 computational performance, the M1 CPU and GPU have similar performance with a peak measured at 1.36 FP32 TFLOPS, while starting from the M2, the GPU significantly outperforms the CPU, i.e., the M4 achieves a peak of 2.9 FP32 TFLOPS while the CPU only achieves 1.5 FP32 TFLOPS. For power consumption, our measurements range from a few to 20 Watts, with similar values for both the CPU and GPU.  

The contributions of this paper are the following:

\begin{itemize}
    \item A comprehensive architectural review of the Apple Silicon M-Series SoCs (M1, M2, M3, and M4), focusing on the CPU and GPU designs, unified memory architecture, and co-processors such as AMX.
    \item An introduction to the programming models for both the CPU and GPU, offering insights into how to effectively utilize these components for computational workloads.
    \item Design and implementation of a STREAM benchmark to assess memory bandwidth between the CPU/GPU and unified memory, and comparison to HPC SOTA.
    \item Design, implementation, and performance measurement of various GEMM versions in FP32 FLOPS, utilizing different frameworks and implementations, such as the Accelerate framework for the CPU, MPS, and various two Metal shaders for the GPU, and comparing these to HPC SOTA.
    \item Development of a benchmark framework for power consumption measurements using Apple's \texttt{powermetrics} tool to evaluate energy efficiency during GEMM computations, providing insight into the performance-per-watt characteristics of the Apple Silicon M-Series.
\end{itemize}

\section{Apple Silicon M-Series SoC Architecture and Programmability}  
Table~\ref{tab:m_series_comparison} compares the architectural features of the Apple Silicon M-Series chips, including the M1, M2, M3, and M4. The table highlights aspects such as process technology, CPU architecture, number of performance and efficiency cores, clock frequency, vector units, cache sizes, neural engine units, and memory technology.
\begin{table*}[htbp]
\centering
\caption{Comparison of Baseline Apple Silicon M Series Architecture.}
\begin{tabular}{@{}p{4.5cm}cccc@{}}
\toprule
\textbf{Feature}            & \textbf{M1} & \textbf{M2} & \textbf{M3} & \textbf{M4} \\ \midrule
\textbf{Process Technology (nm)} & 5          & 5/4       & 3         & 3         \\
\textbf{CPU Architecture}   & ARMv8.5-A   & ARMv8.6-A  & ARMv8.6-A   & ARMv9.2-A   \\
\textbf{Performance/Efficiency Cores} & 4/4        & 4/4       & 4/4    & 4/6    \\
\textbf{Clock Frequency (GHz)} & 3.2 (P)/2.06 (E) & 3.5 (P)/2.42 (E) & 4.05 (P)/2.75 (E) & 4.4 (P)/2.85 (E) \\
\textbf{Vector Unit (name/size)} & NEON/128   & NEON/128  & NEON/128 & NEON/128   \\
\textbf{L1 Cache (KB)}      & 128 (P)/64 (E) & 128 (P)/64 (E) & 128 (P)/64 (E) & 128 (P)/64 (E) \\
\textbf{L2 Cache (MB)}      & 12 (P)/4 (E) & 16 (P)/4 (E) & 16 (P)/4 (E) & 16 (P)/4 (E) \\
\textbf{AMX Characteristics} & FP16,32,64 & FP16,32,64/BF16  & FP16,32,64/BF16 & FP16,32,64/BF16  \\ \midrule
\textbf{GPU Cores}          & 7-8         & 8-10       & 8-10    & 8-10     \\
\textbf{Native Precision Support} & FP32, FP16, INT8 & FP32, FP16, INT8 & FP32, FP16, INT8 & FP32, FP16, INT8 \\ 
\textbf{GPU Clock Frequency (GHz)} & 1.27       & 1.39        & 1.38       & 1.47       \\
\textbf{Theoretical FP32 FLOPS (TFLOPS)} & 2.29-2.61    & 2.86-3.57   & 2.82-3.53   & 4.26\\
\textbf{Neural Engine Units (Core)} & 16          & 16         & 16        & 16     \\  \midrule
\textbf{Memory Technology}  & LPDDR4X     & LPDDR5     & LPDDR5   & LPDDR5X   \\
\textbf{Max Unified Memory (GB)} & 8-16         & 8-16-24         & 8-16-24        &  16-24-32        \\ 
\textbf{Memory Bandwidth (GB/s)} & 67      & 100   & 100      & 120      \\ 
\bottomrule
\end{tabular}
\label{tab:m_series_comparison}
\end{table*}

\subsection{Processor}
The Apple Silicon M-Series processors use ARM-based architectures. These processors use a \texttt{big.LITTLE} approach, which incorporates a mix of high-performance cores (e.g., Firestorm in M1, Avalanche in M2) and efficiency cores (e.g., Icestorm in M1, Blizzard in M2). This architecture enables the processor to dynamically allocate tasks between cores, using performance cores for demanding workloads and efficiency cores for lower-power tasks. The M-Series integrates vector processing capabilities, through NEON 128-bit instructions. The cache hierarchy consists of L1 caches (e.g., 192 KB per performance core) and shared L2 caches of up to 24 MB in the latest M4 chip. Additionally, Apple has integrated the Apple Matrix eXtension (AMX) co-processor, working in conjunction with the CPU. AMX does not execute independently but is controlled via instructions from the CPU. Note that Apple AMX is not Intel AMX (Advanced Matrix Extensions), which is an instruction set extension for x86 processors.

AMX can process multiple matrix elements in parallel. AMX supports fixed matrix dimensions (e.g., 4×4 or 8×8) for its operations, regardless of the size in bytes of the individual elements, such as FP16, FP32, or INT8. AMX processes data in tiles designed to operate on a fixed number of elements. The hardware does not differentiate between precisions at the tile level but adjusts its data fetching and computation pipelines to accommodate the varying byte sizes. The AMX unit keeps evolving while the M1 - M4 iteration, especially in the latest M4, standardized ARM SME (Scalable Matrix Extension) is equipped, which is later proved to be fairly similar to the AMX unit at its core~\cite{remke2024hello}. 

\noindent \textbf{Programmability.} The programmability of Apple Silicon M-Series CPUs is largely based on the Clang/LLVM toolchain, which provides support for the ARM architecture and allows developers to write code in languages like C, Objective-C (a hybrid programming language that combines the features of Objective-C and optionally C++ used within the Apple ecosystem), C++, and Swift. These compilers optimize for the M-Series’ performance and efficiency cores. For programming the CPU's vector units, developers can use ARM intrinsics to write SIMD (Single Instruction Multiple Data) operations explicitly. AMX extends the ARM instruction set to include undocumented matrix-specific operations, which include instructions for loading, processing, and storing matrix data.

For higher-level abstractions, Apple provides the \textit{Accelerate} framework, which includes highly optimized libraries for numerical computing. Among these is vDSP (vector Digital Signal Processing), a part of Accelerate that simplifies the implementation of signal processing and linear algebra tasks while automatically leveraging the vector and AMX capabilities of the CPU for performance gains. BLAS routines within Accelerate also ensure that basic linear algebra operations are optimized for the M-Series architecture, utilizing the AMX vector units for enhanced throughput. 

At an even higher level of abstraction, it is possible to use frameworks, such as Core ML or TensorFlow Lite, to exploit vector instructions, AMX, and the Neural Engine automatically on Apple hardware. This bypasses the need to program vectorized instructions explicitly.

\begin{tcolorbox}[boxrule=0.5pt, colback=gray!20, colframe=black, sharp corners, breakable]
\textbf{HPC Perspective}:
Objective-C (and Objective-C++) applies a set of modifications to C (and C++, respectively). This makes it both low-level and easy to pick up for programmers coming from HPC or generic C/C++ backgrounds, though the unusual syntax for message passing requires some acclimatization. In practice, C/C++ programs can be compiled directly as Objective-C/Objective-C++. Compared to Fortran, Objective-C and Objective-C++ are much more accessible to programmers familiar with systems languages, and are easy to learn ad-lib due to Apple's extensive documentation and examples. The Apple Accelerate framework provides HPC BLAS and LAPACK libraries, capable of using AMX. They are specifically optimized for high performance and low power consumption.
\end{tcolorbox}

\subsection{Graphics Processing Unit} The Apple Silicon M-Series GPUs employ a tile-based deferred rendering (TBDR) architecture.
In the TBDR approach, the scene to be visualized is divided into smaller tiles, and rendering calculations are performed on a tile-by-tile basis. It divides the screen into tiles and processes them sequentially, deferring shading calculations until the visibility of pixels is determined. This minimizes the rendering of hidden pixels, saving processing power. The M-Series GPUs integrate on-chip caches, including per-core and shared memory pools. The design scales across M-Series iterations, with the number of GPU cores increasing from 7-8 in the M1 to 8-10 in the M4, along with improved cache sizes. 

\noindent \textbf{Programmability.} Apple M-Series GPU's primary programming interface is the Metal API, a low-level, high-performance graphics and compute framework. Metal enables fine-grained control over GPU resources and execution, supporting tasks ranging from graphics rendering to general-purpose compute workloads like matrix multiplication and machine learning. It is possible to write GPU code in Metal Shading Language (MSL), similar to C++ or CUDA. The M-Series GPUs' TBDR architecture requires tile-level optimizations. Apple also has higher-level programming frameworks, such as Metal Performance Shaders (MPS), designed to accelerate various computational tasks on the GPU. MPS provided a high-level abstraction for efficiently utilizing the hardware’s capabilities. MPS is built on top of the Metal API, allowing us to access a wide range of pre-optimized GPU operations. For instance, MPS includes optimized implementations for operations, such as matrix multiplication, convolution, and reduction, which are also commonly used in HPC. These shaders automatically utilize the advanced features of M-Series GPUs, such as ray tracing, mesh shading, and dynamic caching, without requiring developers to manage the underlying hardware complexities.
\begin{tcolorbox}[boxrule=0.5pt, colback=gray!20, colframe=black, sharp corners, breakable] 
\textbf{HPC Perspective}: 
The Metal framework provides low-level GPU access, similar to Vulkan and OpenGL on other platforms. It can be used with Objective-C/C++ or Swift, though pure C++ bindings also exist. Its pipeline supports the execution of shaders. Metal Performance Shaders are first-party HPC kernels by Apple, that provide many kernels for HPC tasks such as matrix multiplication. Established GPU software in HPC includes CUDA and OpenCL. Comparably, Metal makes extensive use of kernels and lets programmers write their own. The Metal API shares a lot of similarities with OpenCL, though it is a little higher-level than that of CUDA or cuBLAS. 
\end{tcolorbox}

\subsection{Neural Engine} The Apple Silicon M-Series Neural Engine is a specialized hardware accelerator that optimizes Machine Learning (ML) workload. First introduced with the M1 chip, the Neural Engine features 16 processing units (cores) in the M1, M2, M3, and M4. It is optimized for tensor-based operations, such as matrix multiplications. The Neural Engine supports INT8 and FP16 precision. As a hardware accelerator, the Neural Engine operates independently of the CPU and GPU, not as a co-processor, such as AMX. The Neural Engine delivers higher throughput for matrix operations than AMX but at lower precision in FP16. While this makes it highly efficient for AI-related tasks, low numerical precision is not beneficial for traditional HPC workloads. Consequently, HPC applications requiring FP32 or FP64 precision may not fully benefit from the Neural Engine. 

\noindent \textbf{Programmability.} Core ML allows us to integrate pre-trained machine learning models into applications, automatically optimizing them for potential execution on the Neural Engine. 
However, the low-level programmability of the Neural Engine is extremely limited compared to the Apple Silicon CPU or GPU. Its internal instructions are undocumented and explicit utilization requires reverse engineering.

\begin{tcolorbox}[boxrule=0.5pt, colback=gray!20, colframe=black, sharp corners, breakable] \textbf{HPC Perspective}: The neural engine is akin to Nvidia Tensor cores \cite{markidis2018nvidia} and AMD GPU matrix cores \cite{schieffer2024rise}: they perform tensor multiplication with low precision. While Nvidia and AMD provide WWMA instructions and BLAS to program them, Apple's Neural Engine cannot be instructed directly. Developers are required to use the unified Core ML framework, which does not provide granular control nor guarantees that the Neural Engine is used for execution.
\end{tcolorbox}

\subsection{Unified Memory}
Unified memory in Apple Silicon M-Series chips is a highly integrated architecture where memory is embedded directly within the System-on-Chip (SoC) rather than being treated as separate, off-chip DRAM. This memory is not just a traditional RAM but is tightly coupled with the CPU, GPU, Neural Engine, and other components within the chip, forming a shared, high-bandwidth memory pool. By integrating memory within the SoC, Apple reduces the need for multiple memory controllers, potentially eliminating the performance overhead associated with accessing external RAM. Integrated into the M-Series chip, the memory controller dynamically allocates resources across different compute units. 

\noindent \textbf{Programmability.} For CPU-only allocations, we can use standard \texttt{malloc} or similar system memory allocators. Memory allocated this way is only visible to and directly accessible by the CPU. If the GPU needs to access this memory, developers must explicitly transfer the data or use shared memory mechanisms provided by the Metal framework.

To allocate memory accessible by both the CPU and GPU, the Metal framework provides APIs, such as \newline \texttt{newBufferWithLength:options:} via the \texttt{MTLDevice} object. By specifying \texttt{MTLResourceStorageModeShared}, we can create page-aligned buffers accessible by both the CPU and GPU. This eliminates manual data transfers, as the same physical memory is used. For GPU-specific allocations optimized for high performance, the \texttt{MTLResourceStorageModePrivate} mode can be used, but this memory is not directly accessible by the CPU.
\begin{tcolorbox}[boxrule=0.5pt, colback=gray!20, colframe=black, sharp corners, breakable] \textbf{HPC Perspective}: 
While CUDA also provides high-performance unified memory, data may exist in multiple physical locations (i.e. CPU RAM and GPU VRAM), though the same address space is used and the different devices are abstracted away from the user. This is the same case on AMD GPUs, using the HIP API.

\end{tcolorbox}

\section{Benchmarking Apple Silicon M-Series CPU and GPU} 
To evaluate the performance of the Apple SoC, an empirical methodology is used. This consists of three parts, measuring the bandwidth from/to CPU/GPU to/from unified memory and, measuring GEMM FLOPS and energy dissipation using various implementations.

\subsection{Measuring Memory Bandwidth} \label{sub:measurements}
As the first step of our study, we develop a STREAM benchmark, specifically designed to measure the bandwidth to/from CPU/GPU. The STREAM was initially developed for CPU memory bandwidth benchmarking. In order to uncover the potential of the unified memory architecture in M-series chips, a GPU version STREAM for Apple Silicon M-SERIES is needed. We adopt the STREAM benchmark from a CUDA/HIP GPU version~\cite{gpuSTREAM,schieffer2024harnessing}, ported the \texttt{Copy}, \texttt{Scale}, \texttt{Add}, and \texttt{Triad} kernels with MSL, and implemented the main logic with Objective-C++. 

When it comes to the CPU, the original \texttt{stream.c} by John D. McCalpin is used, which utilizes OpenMP to control the CPU threads used in the benchmark. In this study, every chip model was tested multiple times with \texttt{OMP\_NUM\_THREADS} threads set from one to the number of physical cores for the respective CPUs, to get the maximum reachable CPU bandwidth. The STREAM code used for the CPU and GPU can be found in the project repository~\cite{prjRepo}.

\subsection{Measuring CPU and GPU Computational Performance}
To measure the computational CPU and GPU performance we measure the performance of different implementations of square matrix-matrix multiplications, which are summarized in Table~\ref{tab:implementations}. To calculate the FLOPS, the number of multiplications ($n^2 (2n - 1)$) is divided by the total execution time in seconds.
\begin{table}[ht]
    \centering
    \caption{Overview of matrix multiplication implementations.}    \label{tab:implementations}
    \begin{tabular}{|c|c|c|}
    \hline
    \textbf{Implementation} & \textbf{Framework} & \textbf{Hardware} \\
    \hline
    Naive algorithm & C++ & CPU \\
    BLAS/vDSP & Accelerate & CPU \\
    \hline
    Naive algorithm as shader & Metal & GPU \\
    Cutlass-style tiled shader & Metal & GPU \\
    Metal Performance Shaders (MPS) & Metal & GPU \\
    \hline
    \end{tabular}
\end{table}
To test matrix multiplication performance, we use $n \times n $ square matrices with size powers of two as this provides further hardware optimizations and as padding to such sizes occurs often. The matrices are dense and initialized as \textit{single-precision} ${R}^{n \times n} \in [0, 1]$. The code to generate the matrices is distributed with the source code~\cite{prjRepo}. All matrices (input and output) are allocated via \texttt{aligned\_alloc}, using a page size of 16,384 bytes. Allocation lengths were automatically extended to the nearest page multiple if a matrix was not perfectly divisible by page size, such that the GPU could bypass memory coping.

\noindent \textbf{CPU Implementations.} An implementation of the standard algorithm with a triple nested loop provides a reference baseline. In addition, we use Accelerate SGEMM as shown in Listing~\ref{lst:cpu}. We also use a multi-threaded tiled matrix-matrix multiplication with OpenMP, using an open-source implementation\footnote{\url{https://github.com/dmitrydonchenko/Block-Matrix-Multiplication-OpenMP}}.

\begin{lstlisting}[
    language=c++, 
    caption=Example CPU implementation utilizing the BLAS subroutine. The function is the callback to the test library., 
    label={lst:cpu}, 
    basicstyle=\ttfamily\tiny,  % Reduce font size
    numbers=left, 
    numberstyle=\tiny, 
    commentstyle=\color{green!60!black}, 
    keywordstyle=\color{blue!80!black}, 
    stringstyle=\color{orange}, 
    breaklines=true, 
    frame=tb, 
    tabsize=4, 
    aboveskip=5pt,  % Reduce space above
    belowskip=5pt,  % Reduce space below
    lineskip=-1pt   % Reduce line spacing
]
#include <Accelerate/Accelerate.h>
...
int n = static_cast<int>(nu);
...
cblas_sgemm(CblasRowMajor, CblasNoTrans, CblasNoTrans, n, n, n, 1, left, n, right, n, 0, out, n);
\end{lstlisting}

\noindent\textbf{GPU Implementations.} We use the Metal framework and custom shaders, together with Apple MPS. The naive and tiled shaders (called Cutlass-style) are obtained from an open-source repository~\footnote{\url{https://github.com/bkvogel/metal_performance_testing}} and compiled into a \texttt{.metallib} library. These shaders are then loaded by their respective implementations on startup before the benchmarks. This is unnecessary for MPS, as the shaders are already pre-loaded. During the benchmarks, an MTL-shared no-copy buffer is made to wrap around the matrix data. The thread group counts/sizes needed to be provided for the naive and tiled (but not MPS) shaders --  eight horizontal and eight vertical thread groups were used. The MPS implementation can be found in Listing~\ref{lst:gpu}.

\begin{lstlisting}[
    language=c++, 
    caption=Example GPU implementation utilizing Metal Performance Shaders. Some assertions are removed for brevity., 
    label={lst:gpu}, 
    basicstyle=\ttfamily\tiny,  % Reduce font size
    numbers=left, 
    numberstyle=\tiny, 
    commentstyle=\color{green!60!black}, 
    keywordstyle=\color{blue!80!black}, 
    stringstyle=\color{orange}, 
    breaklines=true, 
    frame=tb, 
    tabsize=4, 
    aboveskip=5pt,  % Reduce space above
    belowskip=5pt,  % Reduce space below
    lineskip=-1pt   % Reduce line spacing
]
#include <MetalPerformanceShaders/MetalPerformanceShaders.h>
...
int main() {
    id<MTLDevice> device = MTLCreateSystemDefaultDevice();
    test_suite([device](unsigned int n, unsigned int memory_length, void* left, void* right, void* out) {
        id<MTLBuffer> bufA = [device newBufferWithBytesNoCopy: left length:memory_length options:MTLResourceStorageModeShared deallocator:nil];
        id<MTLBuffer> bufB = ...
        id<MTLBuffer> bufC = ...
        MPSMatrixDescriptor *desc = [MPSMatrixDescriptor matrixDescriptorWithRows:n columns:n rowBytes:n * sizeof(float) dataType:MPSDataTypeFloat32];
        MPSMatrix *matA = [[MPSMatrix alloc] initWithBuffer:bufA descriptor:desc];
        MPSMatrix *matB = ...
        MPSMatrix *matC = ...
        MPSMatrixMultiplication *matrixMultiplication = [[MPSMatrixMultiplication alloc]
                initWithDevice:device resultRows:n resultColumns:n interiorColumns:n];
        id<MTLCommandQueue> commandQueue = [device newCommandQueue];
        id<MTLCommandBuffer> commandBuffer = [commandQueue commandBuffer];
        [matrixMultiplication encodeToCommandBuffer:commandBuffer leftMatrix:matA rightMatrix:matB resultMatrix:matC];
        [commandBuffer commit];
        [commandBuffer waitUntilCompleted];
    }, "../../../../data/");
    [device release];
}
\end{lstlisting}

\subsection{Measuring Power Consumption} %
For power measurement of matrix-matrix multiplication, we use the first-party utility \texttt{powermetrics}. This provides regular real-time hardware diagnostics, including separate power usage for the CPU and GPU. The monitor starts sampling CPU and GPU power \textit{without} automated sampling: \texttt{powermetrics -i 0 -a 0 -s cpu\_power,gpu\_power -o FILENAME}, waiting for \texttt{SIGINFO} signal to take samples. After two seconds (to ensure the utility is warmed up), a \texttt{SIGINFO} is sent to reset the sampler before the multiplication runs. After the multiplication, the second \texttt{SIGINFO} is sent, thereafter shutting down the monitor. The utility thus \textit{measures total dissipation} between startup/previous signals. This is confirmed empirically while exploring the tool.

\section{Experimental Setup}
Our experiments use four Apple Silicon M-Series processors: M1, M2, M3, and M4, covering all released M-series generations in January 2025. All four chips are equipped with the maximum number of CPU and GPU cores of the \textit{base models}, as described in Table~\ref{tab:m_series_comparison}. Information on the computer models used in this study is listed in Table~\ref{tab:hardware}.

\begin{table}[ht]
    \centering
    \caption{Basic information of devices used.}    \label{tab:hardware}
    \begin{tabular}{|c|c|c|c|c|}
    \hline
    \textbf{Feature} & \textbf{M1} & \textbf{M2} & \textbf{M3} & \textbf{M4} \\
    \hline
    Device & MacBook Air & Mac mini & MacBook Air & Mac mini \\
    \hline
    Release & 2020 & 2023 & 2024 & 2024 \\
    \hline
    Memory & 8GB & \multicolumn{3}{|c|}{16GB}\\
    \hline
    Cooling & Passive & Air & Passive & Air \\
    \hline
    MacOS  & 14.7.2 & 15.1.1 & 15.2 & 15.1.1 \\
    \hline
    \end{tabular}
\end{table}

For software environment details and toolchains, refer to the table in the \texttt{README} document in our repository~\cite{prjRepo}.

For the input matrices, $n \times n $ square matrices are used, with values of $n$ as follows: 32, 64, 128, 256, 512, 1,024, 2,048, 4,096, 8,192, 16,384. All implementations are tested on all these sizes. Except for CPU-Single (Baseline) and CPU-OMP, which did not execute 8,192 and 16,384 due to the long execution time. 

All tests are carried out in a normal indoor environment with the power supply connected to ensure the chip operates at its maximum performance. The computer is kept awake via \texttt{caffeinate}. During the testing process, an effort is made to ensure that no other programs interfere with or consume system resources. Typically, tests are conducted after a system reboot, followed by an idle period until the system is fully idle.

In order to provide reference points, benchmarks are also performed on an internal Nvidia GH200 system. This superchip has 480GB LPDDR5X RAM, with 96GB HBM3 memory. For Grace CPU and Hopper GPU memory bandwidth measurements, the STREAM tests in the official Nvidia HPC benchmark 24.9 are used. For GEMM performance evaluation, the  \texttt{cublasSgemm} in cuBLAS 12.4.2 is used, while both CUDA core and Tensor core (TF32 accelerated path, as FP32 is not supported) performance are tested.

The STREAM benchmark automatically outputs the bandwidth for \texttt{Copy}, \texttt{Scale}, \texttt{Add}, and \texttt{Triad}. The CPU-based STREAM repeated the experiment ten times, and the GPU-based STREAM twenty times. In both cases, only the maximum bandwidth is considered.

The time taken to multiply the matrices is measured by taking the difference between \newline \texttt{std::chrono::high\_resolution\_clock::now()} before and after running the multiplication algorithm, excluding program setup time. The time delta is reported in nanosecond granularity. The matrix-matrix multiplication performs $n^2 (2n - 1)$ floating point operations, from which (G)FLOPS are derived. Each experiment was repeated five times.

The power measurement occurs during the run in which CPU/GPU performance is measured. The results are written into a text file, which is then parsed into a numeric format. Since it piggybacks the computational performance experiment, it too sees five repetitions.

\section{Results} 
In this section, we present different performance results obtained on Apple Silicon M-Series SoCs. We also benchmark an Nvidia GH200 superchip to provide a comparison to HPC's current state of the art.

\subsection{Memory Bandwidth Measurements} 

\begin{figure}
    \centering
    \includegraphics[width=\linewidth]{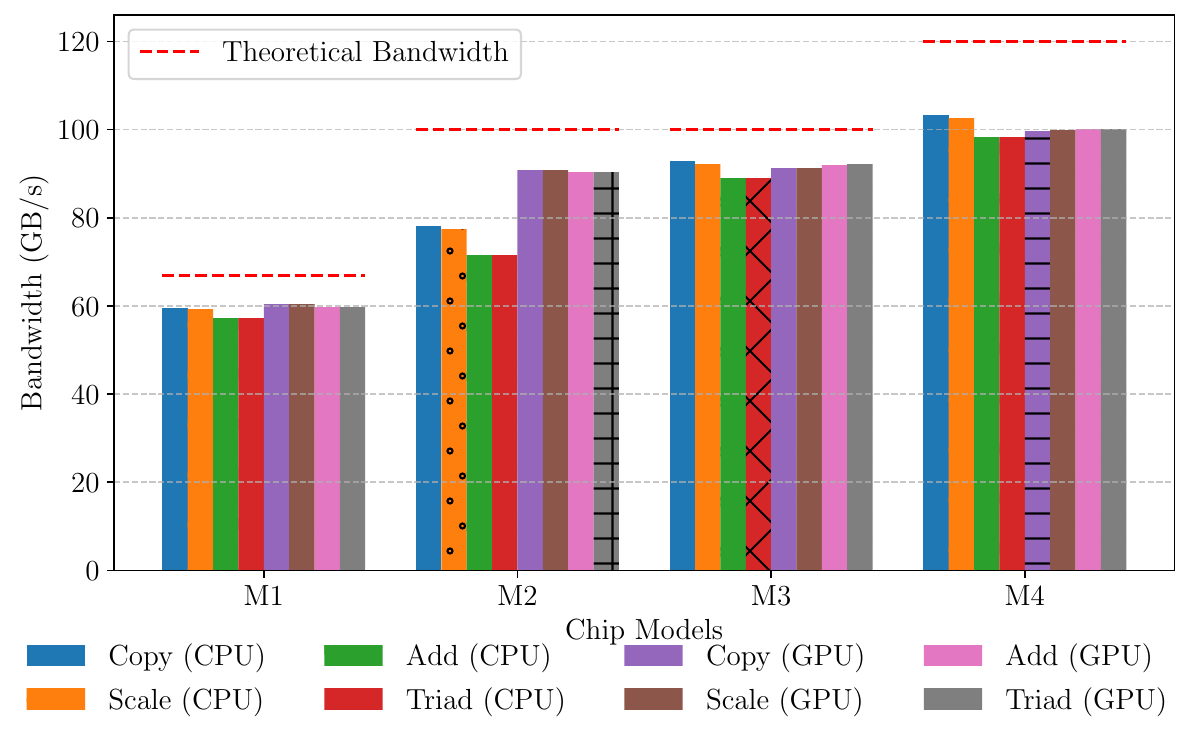}
    \caption{STREAM benchmark results of each processor. The line represents the theoretical bandwidth, and CPU/GPU implementations are shown.}
    \label{fig:stream}
\end{figure}

The STREAM benchmark indicates the memory bandwidth, both CPU and GPU and can be used to compare theoretical and actual bandwidth. Figure~\ref{fig:stream} shows the STREAM benchmark performance for our experiments. M1 to M4 (respectively) see up to 59~GB/s, 78~GB/s, 92~GB/s, and 103~GB/s bandwidth for CPU; 60~GB/s, 91~GB/s, 92~GB/s, and 100~GB/s for GPU. All chips get to $\approx 85\%$ of theoretical peak bandwidth -- within 10~GB/s (M1, M3) and 20~GB/s (M4). The M2 CPU deviates with a 20-30~GB/s gap comparing the Copy and Scale to other kernels, this is about $13\%$. Since the theoretical peaks on M2 and M3 are the same and GPU-based kernels can achieve the same bandwidth on these two chips, CPU-to-memory connectivity is likely less efficient. It is unclear why the M2's CPU performed worse than anticipated.

\begin{tcolorbox}[boxrule=0.5pt, colback=gray!20, colframe=black, sharp corners, breakable] \textbf{HPC Perspective}: Apple Silicon achieves 78~GB/s (78\% of theoretical, M2 CPU) to 92~GB/s\% (M3 CPU, 92\%) bandwidth. In comparison to existing HPC systems, the vastly more capable GH200 attained 310~GB/s (81\%) when using CPU memory, and 3700~GB/s (94\%) using HBM3. In literature, AMD MI250X is observed to reach 85\% of its theoretical peak at only 28 GB/s \cite{schieffer_understanding_2024}. Comparatively, Apple Silicon sees better memory efficiency when using CPU. 
\end{tcolorbox}

\subsection{CPU and GPU Computational Performance} 

\begin{figure*}[htbp] 
    \centering
    \includegraphics[width=\textwidth]{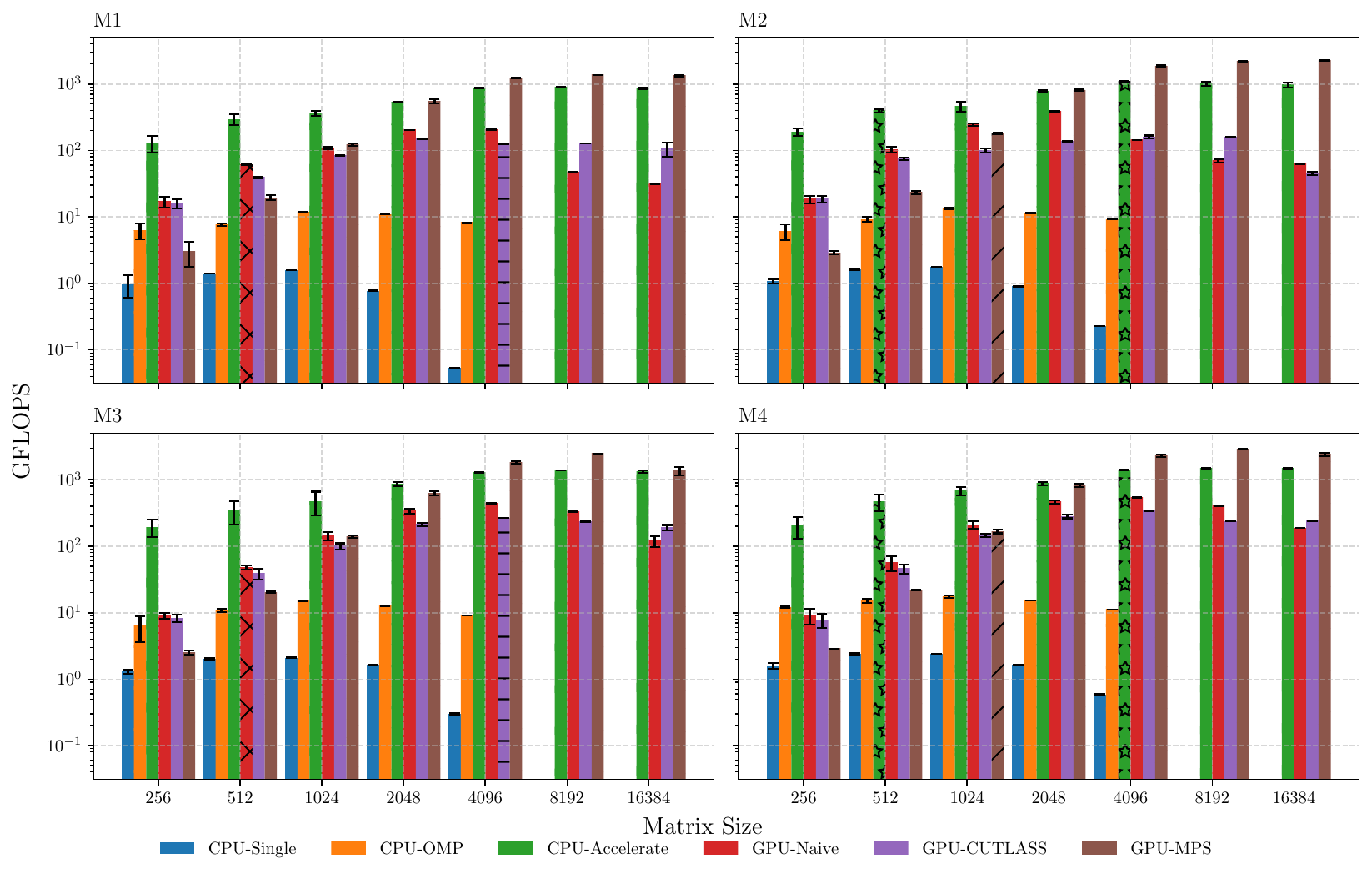} 
    \caption{GFLOPS for all implementations and matrices sizes.}
    \label{fig:GFLOPs-ALL}
\end{figure*}

We evaluate the computational performance of the SoC for both CPU and GPU, in terms of FP32 FLOPS. The results of our GEMM matrix-matrix multiplication performance of the different implementations across M1, M2, M3, and M4,  are shown in Figure~\ref{fig:GFLOPs-ALL}. The vDSP and BLAS implementations perform nearly identically, and thus, only vDSP is considered (listed as ``Accelerate") -- they assumedly both run on AMX. For each processor, we evaluate the floating-point performance across a range of matrix sizes. 

On the CPU, the naive baseline implementation demonstrated the execution time growth of a cubic algorithm.
It consistently underperforms compared to the highly optimized BLAS and vDSP libraries, with vDSP achieving the highest performance 
(0.90~TFLOPS on M1, 1.09~T on M2, 1.38~T on M3, and 1.49~T on M4)
among CPU-based methods. On the GPU, Metal Performance Shaders (MPS) 
dominate (1.36~TFLOPS on M1, 2.24~T on M2, 2.47~T on M3, and 2.9~T on M4) other GPU implementations, with the Metal implementation using Cutlass-style shaders closely following (0.15~TFLOPS on M1, 0.16~T on M2, 0.27~T on M3, and 0.34~T on M4),
while naive shaders lagging in performance (0.2 TFLOPS on M1, 0.39~T on M2, 0.45~T on M3 and 0.54~T on M4).
As expected, GPU-based methods significantly outpace their CPU counterparts for larger matrix sizes due to their high degree of parallelism, though they are less optimal at smaller sizes for their large overhead.

When comparing the maximum performance among M-Series processors and GPUs, MPS demonstrates superior FLOPS on all processors, with M4 achieving the highest performance due to its increased GPU core count and improved architecture. Incremental improvements from M1 to M4 processors are evident, with each generation showing better performance metrics and resource utilization.

\begin{tcolorbox}[boxrule=0.5pt, colback=gray!20, colframe=black, sharp corners, breakable] \textbf{HPC Perspective:} The maximum performance is achieved on GPU with MPS (reaching 63\% of theoretical peak at 2.9 TFLOPS on M4). The maximum performance on the CPU is with Accelerate (vDSP), albeit almost 2x slower than MPS. 

The GH200 running \texttt{cublasSgemm} achieves 41 TFLOPS (61\% of theoretical peak) on CUDA cores, and 338 TFLOPS (69\% of theoretical peak) using Tensor Cores (TF32). It should be noted that the comparison to Tensor Cores is unfair since these use mixed precision and the Apple SoC's equivalent would be the Neural Engine, which is not tested. An Intel Xeon CPU Max 9468 (Sapphire Rapids) achieves 5.7 TFLOPS with \textit{double-precision} matrix multiplication~\cite{siegmann_first_2024}.
\end{tcolorbox}

\subsection{Power Consumption} 
\begin{figure*}[ht]
    \centering
    \includegraphics[width=\linewidth]{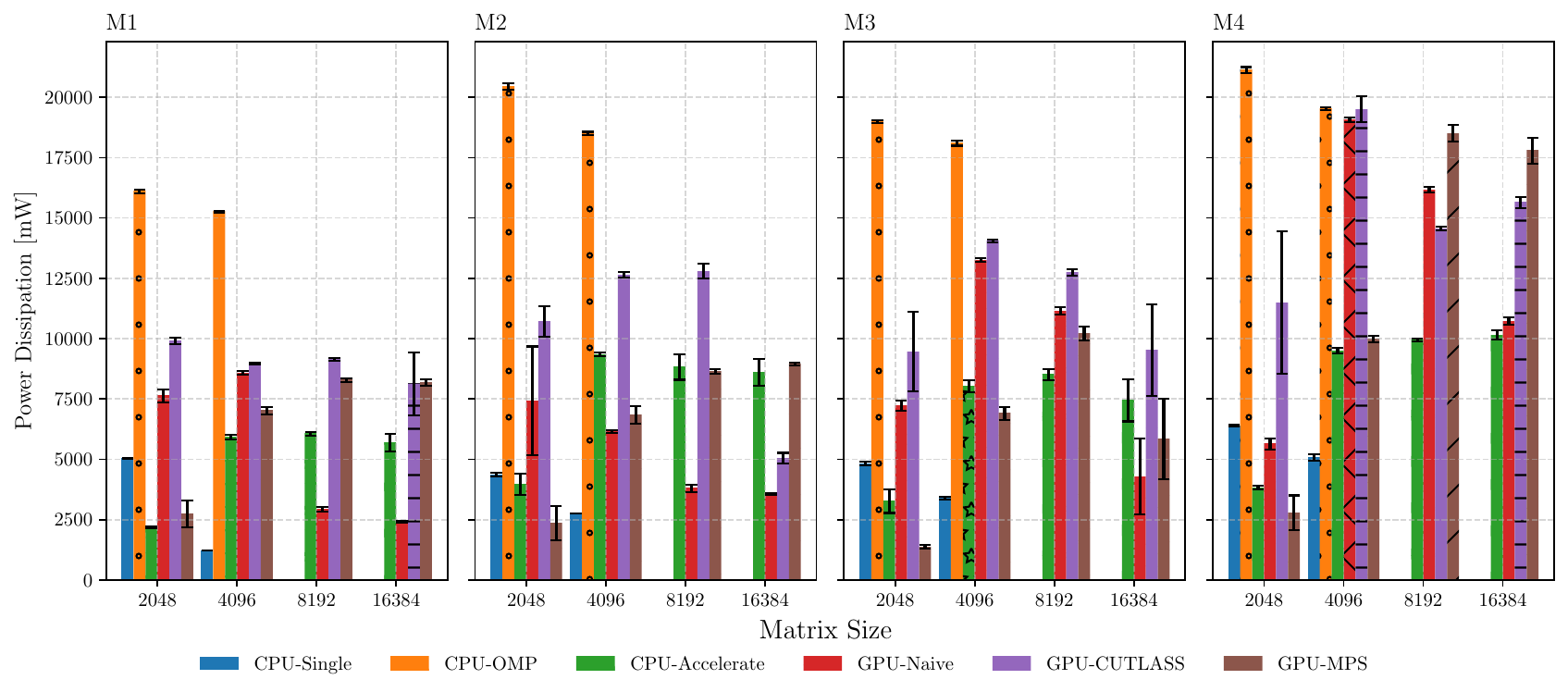}
    \caption{Power utilization of each implementation varying matrix size.}
    \label{fig:power}
\end{figure*}

\begin{figure}[ht]
    \centering
    \includegraphics[width=\linewidth]{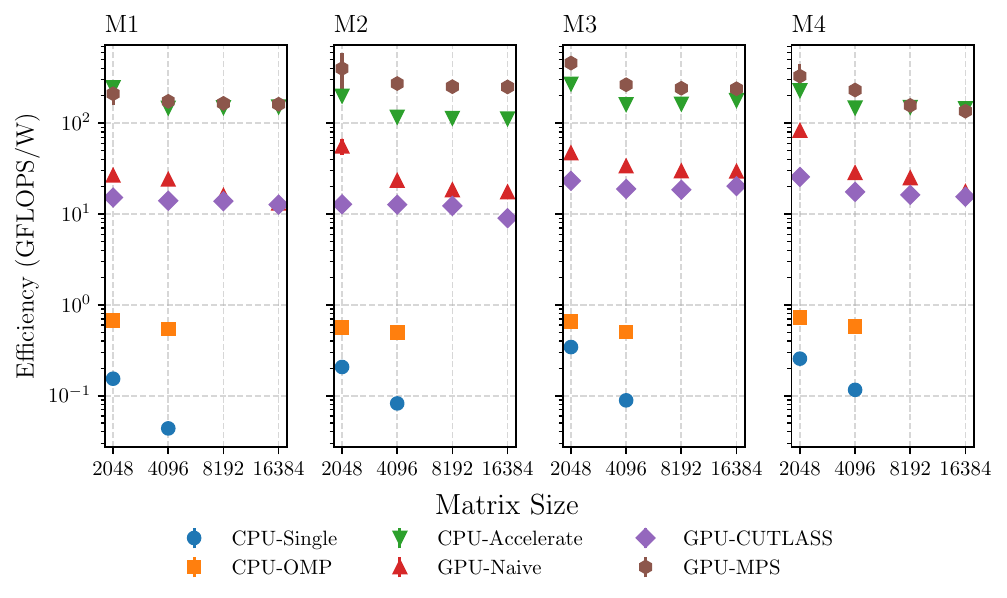} 
    \caption{Power efficiency in GFLOPS per Watt, higher is better.}
    \label{fig:efficiency}
\end{figure}

While running matrix multiplication, we also measure power consumption. The results in Figure~\ref{fig:power} reveal a proportional increase in energy usage with performance across all implementations. M4 exhibited the highest power consumption using the Cutlass-style shader, though on the GPU higher power consumption yields proportionally higher performance. CPU implementations in single and OMP for small problems consume significantly higher power than GPU-based implementations.
When investigating the power efficiency, shown in Figure~\ref{fig:efficiency}, both the GPU and Accelerate accelerated implementations experience similar power dissipation, albeit Apple's first-party Metal Performance Shaders and Accelerate subroutines are by far the most optimized~\ref{fig:efficiency}. All four chips reached the efficiency of 200 GFLOPS per Watt with GPU-MPS (0.21 TFLOPS/W on M1, 0.4 T/W on M2, 0.46 T/W on M3 and 0.33 T/W on M4), about $10\times$ higher efficiency than the other two GPU-based implementations (i.e., GPU-Native and GPU-CUTLASS). CPU-Accelerate also has outstanding power efficiency on A-Series chips (0.25~TFLOPS/W on M1, 0.2~T/W on M2, 0.27~T/W on M3 and 0.23~T/W on M4). The lowest power efficiency is obtained from CPU-based implementations, where both CPU-single and OMP achieve less than 1 GFLOPS per Watt across all four chips. 

\begin{tcolorbox}[boxrule=0.5pt, colback=gray!20, colframe=black, sharp corners, breakable] \textbf{HPC Perspective}: 

\texttt{powermetrics}'s results are software estimates, and Apple explicitly advises against comparing them to different devices.  With this in mind, as a rough indication, the most power-efficient supercomputer on Green500~\cite{top500} runs at 72~GFLOPS/Watt. Our lowest measurement on the M2 CPU achieved 200~GFLOPS/Watt. We were unable to measure power consumption on the GH200 due to time constraints. Literature sees an Nvidia A100 achieve 0.7~TFLOPS per Watt~\cite{luo_benchmarking_2024} using \texttt{mma}, though it should be noted that this is for mixed-precision TensorCores and therefore not a perfectly fair comparison.
\end{tcolorbox}

\section{Related Work} 

\noindent  \textbf{SoC for HPC.} The landscape of HPC has evolved from traditional architectures that rely on separate CPUs and GPUs to heterogeneous SoC designs that integrate multiple processing elements~\cite{kenyon_apple_2022, chien_adaptive_2014, gao_survey_2016}.  This integration facilitates zero-copy memory access, significantly reducing data transfer overheads and enabling more efficient computation~\cite{kenyon_apple_2022}. Previous studies have demonstrated that SoC-based architectures can lead to substantial performance gains in various HPC workloads. For example, optimized zero-copy matrix computations have been shown to reduce execution costs by 34\%~\cite{kato_zero-copy_2013}, while specialized SoC implementations have improved plasma detection accuracy and processing speed~\cite{jablonski_implementation_2021}. \\

\noindent  \textbf{Apple Silicon for HPC Applications.} The introduction of the Apple M1 marked a milestone in ARM-based HPC, influencing other major chip manufacturers to accelerate their ARM SoC development~\cite{liao_impacts_2022}. Although ARM architectures do not inherently provide power efficiency advantages over x86 alternatives~\cite{sankaralingam_detailed_2013}, the M1 chips have demonstrated competitive performance with low power consumption, making them viable for scientific and HPC applications~\cite{kenyon_apple_2022}. Recent studies have explored using Apple Silicon’s unified memory architecture and integrated accelerators for diverse computational workloads. For instance, investigations into machine learning applications have shown performance benefits in training various classifiers~\cite{franco_exploring_2024} and image classification tasks~\cite{kasperek_comparison_2022}. n cryptography, Apple Silicon’s specialized instructions have been utilized to improve execution efficiency and memory footprint in secure computation algorithms~\cite{filho_fast_2024, cryptoeprint:2024/195, zheng_espm-d_2024}.  In addition, previous work in scientific computing -- ranging from real-time rendering~\cite{takeshita_acto3d_2024} and physics simulations \cite{gebraad_seamless_2023} to chemical engineering applications~\cite{cho_performance_2024} -- indicates promising Apple Silicon SoC energy-efficient computation capabilities.

\section{Discussion \& Conclusion}
This paper provides a detailed architectural review of the Apple Silicon M-Series SoCs (M1, M2, M3, and M4), focusing on their CPU and GPU designs, unified memory architecture, and coprocessors such as AMX and SME. Via the design and implementation of STREAM and GEMM benchmarks, we measured computational performance in terms of FP32 FLOPS and explored the power consumption and efficiency characteristics of these chips.

We showed that the M-Series chips demonstrate relatively high memory bandwidth performance, with the unified memory architecture offering comparable bandwidth between the CPU and GPU, closely matching the theoretical values. In comparison, a state-of-the-art Nvidia GH200 achieves similar efficiencies at two orders of magnitude better performance.

The FP32 computational performance of the M1, M2, M3, and M4 chips exhibited improvements throughout the series, with the M4 reaching a peak of 2.9 FP32 TFLOPS. While the M1 showed a peak FP32 performance of 1.36 FP32 TFLOPS, the GPU's performance started to significantly outperform the CPU from the M2 onwards, highlighting the increasing role of the GPU in the M-Series. In comparison, an Nividia GH200 achieves 41 TFLOPS on CUDA cores. 

Additionally, our power consumption measurements showed a range from a few watts to about 20 watts, with similar values for both the CPU and GPU, indicating the potential of the M-Series chips for energy-efficient HPC applications. For comparison, an RTX 4090 performing dense matrix multiplication was found to consume 174~W while reaching 0.51~TFLOPS/W tensor core performance (albeit in MMA, not SGEMM) \cite{luo_benchmarking_2024}, while the M4 reaches 2.9 TFLOPS dissipating only around 20~W and M3 reaches 0.46 TFLOPS/W efficiency with GPU-MPS. While the RTX 4090 outperforms in performance, the overall total footprint of the Apple Silicons remains impressive. We also noticed that the Apple laptops with M1, and M3 SoCs have relatively lower Power Dissipation compared to desktops (M2, M4), which might show the impact of power strategy and cooling methods of different device models. In comparison, the topmost Green500 computer runs at 72 GFLOPS/W, and an Nvidia A100 at 0.7 TFLOPS/W, albeit this comparison is not perfectly fair as it uses TensorCores. 

However, our study has limitations. First, measuring power consumption proved to be challenging. While we used Apple's \texttt{powermetrics} tool to assess energy efficiency, the tool’s measurements may not provide exact values, and discrepancies could arise due to variations in system behavior during computations. Further improvement of power measurement techniques would be necessary to gain a more accurate understanding of the power dynamics in these chips. Moreover, while the M-Series CPUs support a wide range of precisions, including FP64, the GPUs lack native FP64 support. This might limit their suitability for certain scientific applications requiring double-precision arithmetic. We focused on FP32 computations in our analysis, but future studies could explore the impact of mixed-precision workloads on computational efficiency and accuracy.

Future work in this area could explore the performance of the M-Series chips in multi-node or distributed HPC systems. A large gap left behind in this research is the lack of Neural Engine testing, which would better contextualize the M-Series with respect to TensorCore performance. Further investigation on power efficiency and thermal management, particularly in large-scale deployments, would be beneficial in understanding the viability of the M-Series for energy-efficient HPC.

In conclusion, the Apple Silicon M-Series SoCs present a promising platform for HPC workloads, with significant improvements in computational performance and energy efficiency compared to traditional discrete component architectures. While there are some limitations in terms of precision support and power measurement accuracy, we showed that the Apple Silicon M-Series has a promising potential for high-performance, energy-efficient computing. While not a direct replacement for traditional HPC architectures, the M-Series chips offer a different balance of performance and energy efficiency.

\bibliographystyle{AppleSiliconMseriesPaper.bbl}
\bibliography{AppleSiliconMseriesPaper.bbl}

\end{document}